# GMM CLUSTERING FOR IN-DEPTH FOOD ACCESSIBILITY PATTERN EXPLORATION AND PREDICTION MODEL OF FOOD DEMAND BEHAVIOR

*Rahul Srinivas Sucharitha, Seokcheon Lee*

## Abstract

**Understanding the dynamics of food banks' demand from food insecurity is essential in optimizing operational costs and equitable distribution of food, especially when demand is uncertain. Hence, Gaussian Mixture Model (GMM) clustering is selected to extract patterns. The novelty is that GMM clustering is applied to identify the possible causes of food insecurity in a given region, understanding the characteristics and structure of the food assistance network in a particular region, and the clustering result is further utilized to explore the patterns of uncertain food demand behavior and its significant importance in inventory management and redistribution of surplus food thereby developing a two-stage hybrid food demand estimation model. Data obtained from a food bank network in Cleveland, Ohio, is used, and the clusters developed are studied and visualized. The results reveal that this proposed framework can make an in-depth identification of food accessibility and assistance patterns and provides better prediction accuracies of the leveraged statistical and machine learning algorithms by utilizing the GMM clustering results. Also, implementing the proposed framework for case studies based on different levels of planning led to practical results with remarkable ease and comfort intended for the respective planning team.**

# 1. Introduction

Food insecurity implies the uncertainty or absence of the ability to acquire nutritionally satisfactory and safe nourishment in socially acceptable ways (e.g., without resorting to stealing, rummaging, or different adapting strategies). This condition is unavoidable, influencing masses everywhere throughout the world. In the United States, it impacts every community with food insecurity existing in each region in America (Feeding America, 2015). The United States Department of Agriculture (USDA) has evaluated that starting in 2017, nearly 40 million individuals have been living in sustenance-uncertain household units, with 6.5 million of them forming minors (Feeding America, 2015). In the United States, there is a range of assistances connecting collective efforts between government, public and private bodies for food-insecure individuals. One of the largest national non-profit hunger relief organizations tackling hunger and food insecurity is Feeding America.

There are around 200 food banks and close to 60,000 food agencies (Consisting of food pantries and meal programs) that Feeding America is spearheading to offer food and assistance to the food insecure people and households. Food and donations are mostly recovered from national food and grocery producers, suppliers, shippers, packers, and farmers, as well as public bodies and other organizations, and transported to food banks, which serve as food storage and delivery depots for smaller front-line food agencies (Feeding America, n.d.). Hence, the underprivileged population can receive donated foods from these food agencies. In general, in this donated food supply chain network, there is high uncertainty in the supply and demand of donated foods for the food insecure. Food banks assess effectiveness and equity at the individual population level, and their primary customers are the food agencies. Hence, demand at the individual level is usually hidden to the food bank as previously, the charitable food agencies have been providing little to sometimes conflicting information to the food banks (Orgut et al., 2016b). Hence, the challenge of estimating demand is a significant source of uncertainty in non-profit food distribution. As a result of the poor visibility of the actual "need" for the food distribution, food banks frequently overestimate their demand (e.g., taking responsibility for the entire poverty population in a region). They sometimes underestimate demand (e.g., ignoring populations above the poverty level but have hunger needs). Hence, this paper develops and investigates a novel methodology for better estimating the demand with more resistance to demand fluctuations.

The final goal of a bequest-driven supply chain such as the food bank supply chain is to maximize the relief for the people in need while minimizing food waste as a by-product benefit. We fill this gap by explicitly studying the nature of the families visiting the food agencies based on demographical information and define food assistance deserts in the given region of study by applying concepts of unsupervised machine learning (Gaussian Mixture Models (GMM) clustering) to observe the geographic and demographic intricacies of the given region in detail. Once the dataset is examined and analyzed comprehensively, the results obtained from clustering the dataset will be used for forecasting demand-side inputs using supervised machine learning forecasting techniques. Several studies in the public policy and health literature examine the usage of food banks, the challenges associated with limited and unpredictable supply, and the forecasting of supply uncertainty (Nair et al., 2017a). However, to the best of our knowledge, applying statistical analysis techniques to handle the demand uncertainty in the food bank supply chain system has not been addressed. Therefore, we study the nature of the current food-insecure household situation in the given region using unsupervised machine learning methods such as clustering and demonstrate that we can get reasonable estimates for food demand by implementing supervised machine learning techniques on the clustered data. Our results generate forecast accuracy of 82% for specific instances.

Our study has particular merit because it is essential for non-profit organizations to leverage knowledge and technology to renovate and reinvent their preparedness and effectiveness. Food banks having information on their current and future demand behavior can help improve their food distribution efficiency and make suitably informed distribution decisions, thereby meeting their objectives of ensuring adequate, equitable, and efficient food-aid operations and distributions to the people in need. *Equity* implies serving the customer's needs fairly, while *effectiveness* implies the ability to meet the customer's needs, and *efficiency* implies "the inverse of the cost of making and delivering a product to the customer" (Chopra and Meindl, 2021). These costs drive the resources needed to collect and distribute the donated food in food banks.

The remainder of the paper is organized as follows- Section 2 provides a summary of the literature. Section 3 presents the data and methods, and Section 4 summarizes our results, Section 5 provides the discussion by means of case studies, followed by our conclusions in Section 6.

## 2. Literature Review

Food insecurity and hunger are long-term humanitarian issues, requiring the need to consider the necessity for evenhanded dispersion of resources (Orgut et al., 2018). There has been broad research done in humanitarian logistics with significance towards the issues and challenges faced by non-profit food assistance programs such as food banks and food pantries, as enlightened by Davis et al. (2014). Food bank supply chains line up with the description of humanitarian supply chains by responding to the disaster of food insecurity which can occur unexpectedly (i.e., job loss, natural calamity, etc.) or slowly (i.e., destitution) (Balcik et al., 2008). The research presented in this paper aids in finding the different possible factors affecting food insecurity and hunger, thereby facilitating the accessibility of food and equitable distribution of resources to the people in need. According to Waity (2016), there are innumerable ways of studying food accessibility for people. One of these methods is considering food deserts. Food deserts are regions deficient in sources of healthy nutrients. Our paper implements this concept of food accessibility.

There has been a decent amount of work that studies the subject of food bank supply chain. Mathematical models were presented by (Orgut et al., 2016a; Orgut et al., 2017) to enable the equitable and effective distribution of food donations to the people in need. Linear programming models were formulated with the maximization of effectiveness and an equity constraint developed to solve the distribution of donated foods. Deterministic network-flow models were used to reduce the quantity of undistributed food. Several logistical issues that are being faced by the food banks have also been taken into consideration by considering the transportation schedules and permitting food banks to gather food from the limited food donors and finally transporting it to the food agencies (Davis et al., 2014). In their paper, Food Delivery Points (FDPs) were proposed. FDPs were obtained by locating them using geographical information. The vehicle capacity and food degeneration constraints were considered during the assignment of food agencies to the respective FDPs. Using the optimal assignment, schedules were created that reflects the collection and distribution of donated food. Islam and Ivy (2021) developed an assignment and distribution model to identify the efficient and effective assignment of counties to food banks and maximized the food distribution at minimum transportation costs. However, these mathematical models do not investigate the varying demands of the various food agencies from where the accessibility of food is studied. Additionally, the challenges taken forth in these papers, are rooted in the lack of accurate knowledge of the demand the individual food agencies and is largely taken as assumptions. Hence, our investigation in demand uncertainty is critical to manage the decision-making at various levels of the non-profit food supply chain.

Previous non-profit-based supply chain literature considers food demand from the food pantries and other food agencies as a deterministic value. In hindsight, the demand for food is dynamic and uncertain. Food agencies assigned to food banks need to obtain a way to observe the demand for food to understand the possible issues arising out of food insecurity. According to Balcik et al. (2008), demand comes in the form of supplies and people for non-profit organizations and products and services for for-profit organizations, with varying demand patterns for both. Sucharitha and Lee (2018) developed a food distribution policy using suitable welfare and poverty indices and functions to ensure an equitable and fair distribution of donated foods per the people's varying demands and requirements. Their simulation study did not consider the factors that caused demand or food insecurity, relying on several assumptions. Also, the model developed was suitable only for a single-day period. Supply of the donated food can be done based on suitable forecasting procedures. Supply based on this kind of non-profit supply chain would mainly deal with guaranteeing enough inventory for the demand and reviewing the changing nature of the supply of the different types of donated foods.

Regarding implementing suitable data mining techniques, there have been relevant literature discussing the role of these techniques in estimating future supply using historical data in various domains. Using forecasting techniques to estimate the food donation and distribution process dynamics, Davis et al. (2016) performed comprehensive numerical studies to quantify the extent of uncertainty in terms of the food donors, the food products, and the supply chain structure. The number of in-kind donations was estimated using several predictive models. Predictive modeling techniques like multiple linear regression, structural equation modeling, and neural networks were used in Nair et al. (2017) to study the dynamics of food donation behavior thereby, predicting the average daily food donated by different food providers in the given region. Paul and Davis (2021) proposed a predictive ensemble model to forecast the contribution of different donor clusters and showcased the necessary behavioral attributes to classify donors as a result. However, the lack of statistical analysis techniques used to study the demand dynamics in the food bank supply chain has been evident and mentioned as a significant challenge from the non-profit supply chain perspective (Orgut et al., 2016b). Recent work addressing this issue has aided in a better understanding of the mitigation of food insecurity. Alotaik et al. (2017) implemented K-means clustering to identify the food assistance deserts, a term Waity (2016) coined while analyzing the spatial inequality between rural and urban areas in access to food agencies. The results obtained from Alotaik et al.'s (2017) analysis helped target the underserved areas in the given region. Finding trends and detailed observations is possible using unsupervised learning methods such as clustering, which is not the case when implementing spatial analysis (Waity, 2016). However, considering the dataset used consists of variables of different sizes and densities, the affected families and certain traits could have been hidden and unobserved, keeping in mind the lack of flexibility in a clustering technique such as K-means clustering (Verma et al., 2012). Implementing a soft clustering method such as GMM in our study ensures better visibility of traits and hidden features in a dataset featuring spatial and demographic information than K-means clustering (Wang et al., 2019), thereby furthering their effectiveness, goals equity, and efficiency. Our research develops a two-stage hybrid food demand estimation model to identify food accessibility and assistance patterns and provide better demand prediction accuracies. To the best of our knowledge, the open literature does not address the demand behavior within food bank organizations.

In this paper, we address the issue of food insecurity in a given region of Ohio by analyzing the food agency service data provided by Greater Cleveland Food Bank (GCFB). Combining their service data and the demographic data provided by the USDA and implementing the GMM clustering method to the combined data based on the distances between the family visiting the food agency and the food agency serving them, we observe the factors leading to food insecurity and provide ways to increase the accessibility of food. We then implement the clustering results for the food demand predictions by developing predictive models using various statistical learning methods for the dataset modified to contain the number of people visiting the food agency as the response variable. Finally, we assessed the model's performances, both with clustering results and without clustering results, based on their predictive accuracy to select the best model based on both generalizability and ability to capture the data structure.

## 3. Data

### 3.1 Data Description

Greater Cleveland Food Bank (GCFB) provided service data of all the food agencies that obtain food from the distribution methods. Fig.1 shows the study area with the food agencies depicted in red and the families visiting these agencies depicted in grey. The plot was developed based on the dataset values of the latitude and longitude variables of the food agencies and the families visiting these organizations. GCFB distributes to food agencies situated in around five counties in Ohio. The service data consists of 600,000+ records for 2018, where each row represents one service to one family.

It includes each family's latitudinal and longitudinal location points and the represented food agency they have visited during that time. It was imperative to obtain the distance between each family and their visited food agency and to locate them and observe if they were at an acceptable distance or not. The dataset entailed the census tract and census block details of each family. Hence, along with the census details, the distance between the family and the food agency is also aligned and tallied. Since the region under study is predominantly an urban area, we consider a 1-mile demarcation as the threshold level of distance as a food assistance desert or low food accessibility (Mattogno et al., 2014). The GCFB service dataset also provided specifics of the number of children, seniors, and adults in every family serving their food agencies. USDA also provides information on the counties' household income and median household income at the census tract level. We obtained this data to study the observation region's low- and high-income population and aggregated it to the service dataset (see Table 1).

There has not been any available study to estimate food demand in the food bank network for the predictive modeling study. Therefore, to develop this study, the aggregated dataset compiled for the clustering study undergoes data wrangling (Xu & Tian, 2015) to obtain the response variable of the total number of people visiting a particular food agency along with the timestamps and geographical information of the person as input variables for the predictive modeling. The wrangled dataset consists of 15000+ records for the fiscal year of 2017-2018. By doing this, we can predict the number of people visiting a food agency and obtain the amount of food that a food agency would require to satisfy the demands of the clients on a daily, weekly, or monthly basis. For this study, we consider daily demand. However, the dataset can be easily modified to study the weekly and monthly food demand.

We will define the former dataset developed for clustering analysis as the aggregated dataset and the dataset obtained from wrangling the aggregated dataset as the data of food demand (see Table 2) from now on.

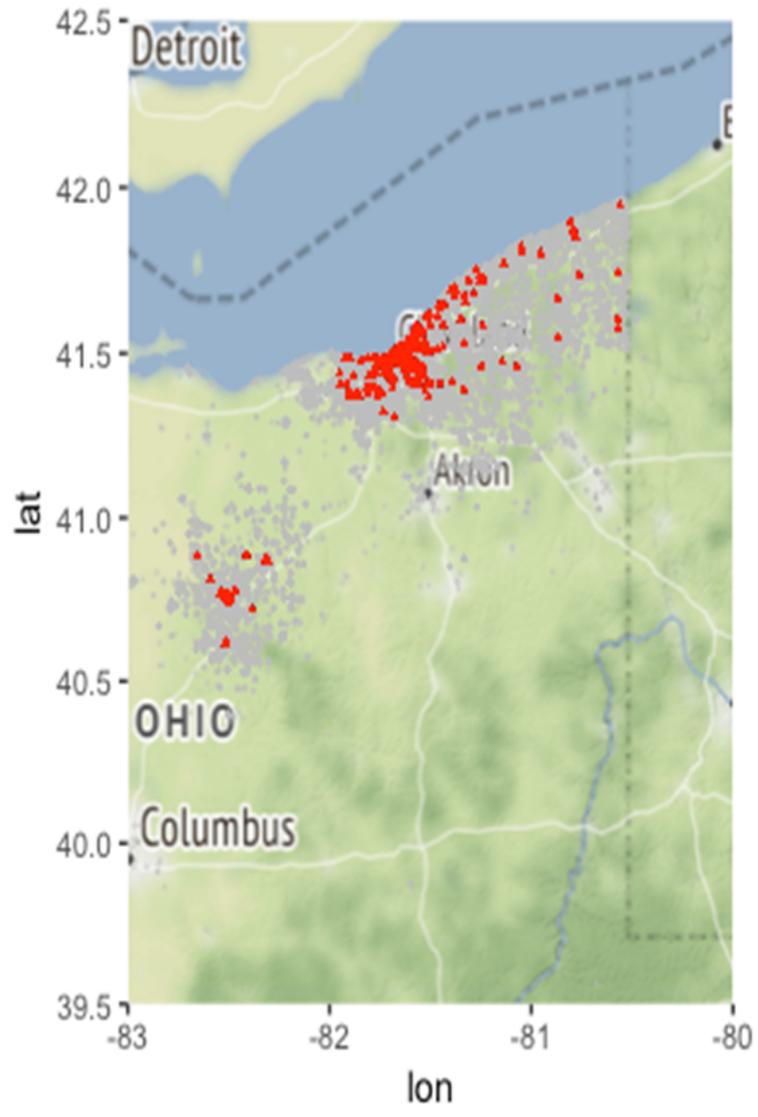

Figure 1 Region of Ohio (Study Area)

Table 1 Aggregated dataset

| Attribute | Description |
|---|---|
| Date | The date that the service took place on |
| Family ID | System generated ID number for each client |
| City | Client's city |
| State | Client's state |
| Zip | Client's zip |
| County | Client's county |
| Count Adult | Number of family members between ages of 18-59 |
| Count Child | Number of family members below age 18 |
| Count Senior | Number of family members age 60 or above |
| Agency Number | ID of the food agency |
| Family Latitude | Client's latitude, obfuscated to three decimal places |
| Family Longitude | Client's longitude, obfuscated to three decimal places |
| Agency Latitude | latitude of agency/pantry |
| Agency Longitude | longitude of agency/pantry |
| County Income | Household income (median) in a given county |

Table 2 Food demand dataset

| Variable names | Description |
|---|---|
| Dow | Day of the week |
| Woy | Week of the year |
| Doy | Day of the year |
| Moy | Month of the year |
| Agency Number | ID of the food agency |
| Food Demand (No. of people) | Total number of people visiting the food agency |

## 3.2 Exploratory analysis

We provide the summary of the descriptive statistics of the food demand dataset by means of the number of people visiting the food agency on a daily basis in Table 3. The table shows that 75% of the overall demand has its value below 108 counts (number of people) while the maximum count is 5588. From observing the distribution of the response variable in Fig. 2, we see the heavy tail of the response variable. The empirical cumulative distribution function plot in Fig. 3 suggests that a small fraction of the response variable includes many people visiting the respective food agency. The y-axis shows the cumulative probability, and the density plots for both datasets are steep and centered at zero, showing that significant events are sporadic and minor events are frequent. This implies we need to estimate the demand by considering several modeling techniques to study the data and not just focus on a one-size-fits-all approach.

Table 3 Summary of Food Demand (Response Variable)

| Parameter | Food Demand (Number of People) |
|---|---|
| **Minimum** | 1.0 |
| **1st Quartile** | 19.0 |
| **Median** | 51.0 |
| **Mean** | 114.9 |
| **3rd Quartile** | 108.0 |
| **Maximum** | 5588.0 |

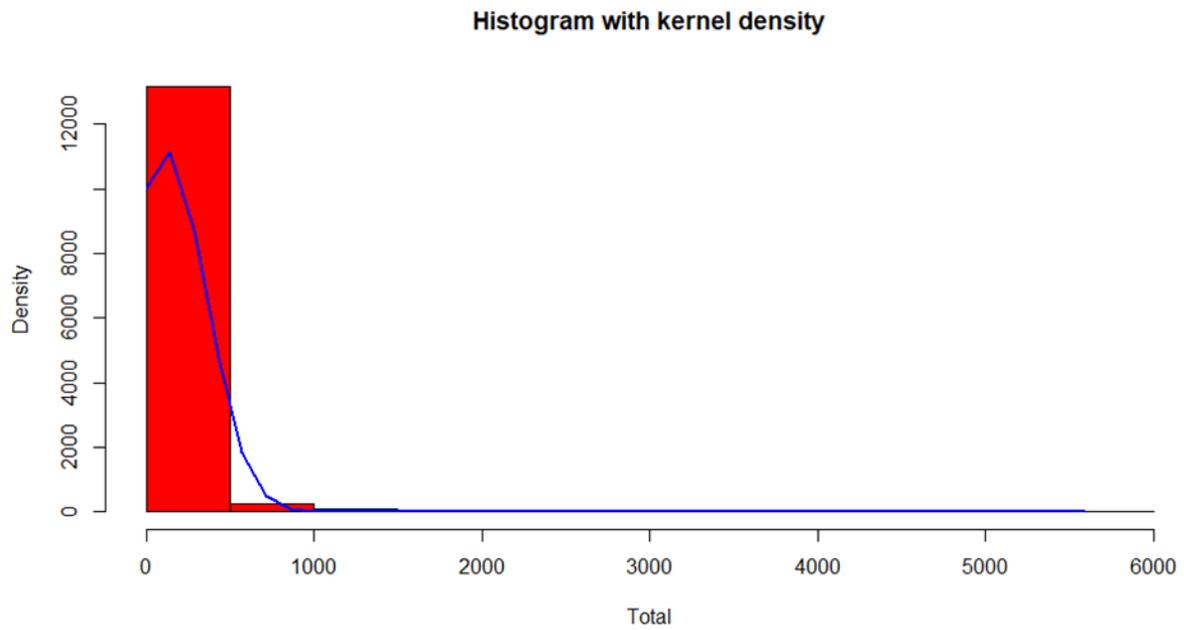

Figure 2 Distribution of response variable: histogram with overlain kernel density plot

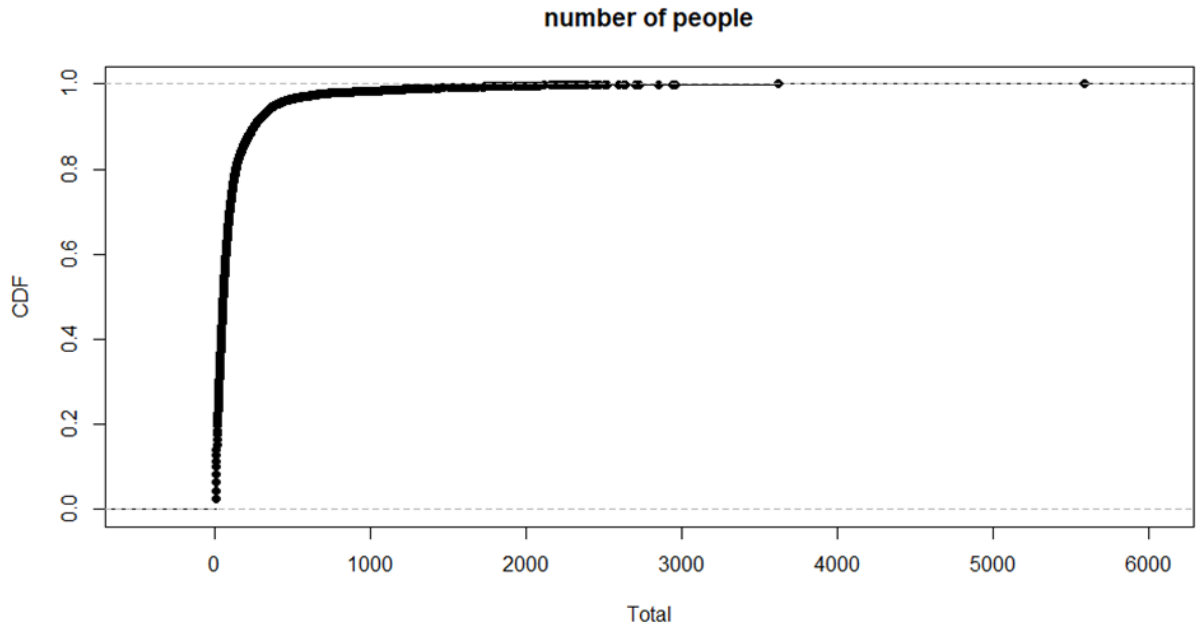

Figure 3 Empirical cumulative distribution functions for the response variable

## 3.3 Methodology

### 3.3.1 Methods

**GMM clustering algorithm for comprehensive data characterization**

The aggregated data involves the data compilation of both the GCFB dataset and the USDA income-related dataset. The distance between each census tract and the assigned food agency and the distance between each family to the assigned food agency will be calculated and saved a variable. After this step, clustering using GMM method is done based on their distances and the demographics mentioned as variables.

The GMM is a common soft clustering method that can approximate any probability distribution by training several weighted variations of Gaussian distributions and thus increasing the number of mixture components. Each gaussian model in our analysis can be thought of as a coverage class. Here, $Y = \left[ Y_1, Y_2, Y_3, ..., Y_d \right]^T$, is denoted as an observation vector where, $Y_1$ is taken as a particular attribute in the given aggregated dataset, and the others are the driver factors. So, $d$ is the number of observation vectors (Ma et al. 2014). The description of GMM is given as follows:

$$p(\theta) = \sum_{k=1}^{K} \alpha_k \phi_k \left( Y | \mu_k \sum_k \right)$$

$$\phi_k\left(Y \middle| \mu_{k'}, \mathring{a}_k\right) = (2\pi)^{-\left(\frac{d}{2}\right)} \left| \mathring{a}_k \right|^{-\left(\frac{1}{2}\right)} \exp exp\left\{ -\frac{1}{2}\left(Y - \mu_k\right)^T \mathring{a}_k^{-1}\left(Y - \mu_k\right)\right\}$$

Where $k$ is the number of mixture models, $\alpha_k$ is the mixture weight with $0 < \alpha_k < 1$, and $\sum_{k=1}^{K} \alpha_k = 1$, $\phi_k\left(\mu_{k'}, \mathring{a}_k\right)$ is the Gaussian model of the $k$th mixture component, $\mu_k$ and $\mathring{a}_k$ denote the mean and the covariance matrix respectively.

Table 4 The geometric characteristics of the basic Gaussian models

| Model | Distribution | Volume | Shape | Orientation |
|-------|-------------|--------|-------|-------------|
| EII | Spherical | Equal | Equal | - |
| VII | Spherical | Variable | Equal | - |
| EEI | Diagonal | Equal | Equal | Coordinate axes |
| VEI | Diagonal | Variable | Equal | Coordinate axes |
| EVI | Diagonal | Equal | Variable | Coordinate axes |
| VVI | Diagonal | Variable | Variable | Coordinate axes |
| EEE | Ellipsoidal | Equal | Equal | Equal |
| EVE | Ellipsoidal | Equal | Variable | Equal |
| VEE | Ellipsoidal | Variable | Equal | Equal |
| VVE | Ellipsoidal | Variable | Variable | Equal |
| EEV | Ellipsoidal | Equal | Equal | Variable |
| VEV | Ellipsoidal | Variable | Equal | Variable |
| EVV | Ellipsoidal | Equal | Variable | Variable |
| VVV | Ellipsoidal | Variable | Variable | Variable |

Each Gaussian model represents a cluster, and the 14 models proposed (Fraley & Raftery, 2007; "Mathematical Statistics and Data Analysis - John A. Rice - Google Books", n.d.) are shown in Table 4. Hence the parameter set of a GMM is composed of $\{\alpha_k, \mu_k, \Sigma_k\}$, with $1 \le k \le K$. The parameters are estimated in the maximum likelihood setting. The optimization is usually carried out using the Expectation-Maximization (EM) algorithms, which are depicted as follows in two steps as follows:

Expectation Step: from the below equation, a posteriori probability $\gamma_{jk}$ at the $j$th data value ($j \in [1, N]$, $N$ denotes the number of samples) is computed based on the randomly given initial values of $\{\alpha_k, \mu_k, \Sigma_k\}$ :

$$\gamma_{jk} = \frac{\alpha_k \phi_k\left(\mu_{k'}, \Sigma_k\right)}{\sum_{k=1}^{K} \alpha_k \phi_k\left(X \middle| \mu_{k'}, \Sigma_k\right)}$$

Maximization Step: In this stage, new set of values for the parameters $\{\alpha_k, \mu_k, \Sigma_k\}$ can be obtained with $\gamma_{jk}$ designed from the earlier mentioned Expectation Step as:

$$\alpha_k = \frac{\phi_k}{N}$$

$$\mu_k = \frac{1}{\phi_k} \sum_{j=1}^{N} \gamma_{jk} X_n$$

$$\sum_k = \frac{1}{\phi_k} \sum_{j=1}^{N} \gamma_{jk} \left( X_n - \mu_k \right) \left( X_n - \mu_k \right)^T$$

Where $\phi_k = \sum_{j=1}^{N} \gamma_{jk}$. A fresh set of $\{\alpha_k, \mu_k, \Sigma_k\}$ can be obtained by Maximization Step, which is applied to the earlier Expectation Step to obtain the new $\gamma_{jk}$. Both these steps are hence iteratively calculated until convergence is obtained based on the likelihood function:

$$L(X) = \sum_{i=1}^{N} ln \left\{ \sum_{k=1}^{K} \alpha_k \phi_k \left( Y | \mu_k \sum_k \right) \right\}$$

Under the pre-set $K$, the final dataset of $\left\{\alpha_k, \mu_k, \Sigma_k\right\}$ is calculated by the maximum $L(X)$, and each $\phi_k \left( Y | \mu_k \sum_k \right)$ is termed a cluster. To obtain the optimal number of clusters ($K$), the following methods are considered.

**Optimal number of Clusters**

A key role in the GMM clustering method is selecting the standard Gaussian model (Table 4) from the 14 types of basic models proposed. One model is chosen at each clustering point, and the data distribution is explained. The best basic model helps to achieve the best clustering performance. Since two or more variables may have a positive or negative correlation, the orientation of the covariances is constrained to be variable across classes. As a result, the EEV, EEE, EVI, and EII models were chosen for the GMM clustering analysis.

Another important function in the GMM clustering approach is to determine the optimal number of components ($K$), which can be obtained through two techniques – The Bayesian information criterion (BIC) (Srivastav, Tewari, and Dong 2013) and the Silhouette score (Rousseeuw, 1987). For BIC, the criterion is formulated as follows:

$$BIC = L(X) - \frac{M}{2} \log \log (N)$$

Where $N$ is the number of samples. The total number of free parameters is represented by $M$ and this is obtained as below:

$$M = Kd + \frac{1}{2}Kd\,(d + 1) + (K - 1)$$

Where $d$ is the number of observation vectors. The $K$ value and the most suitable model that maximizes the BIC typically represents the optimal $(K)$ for the model. Fig. 5 provides the results of the BIC values obtained for the chosen gaussian models.

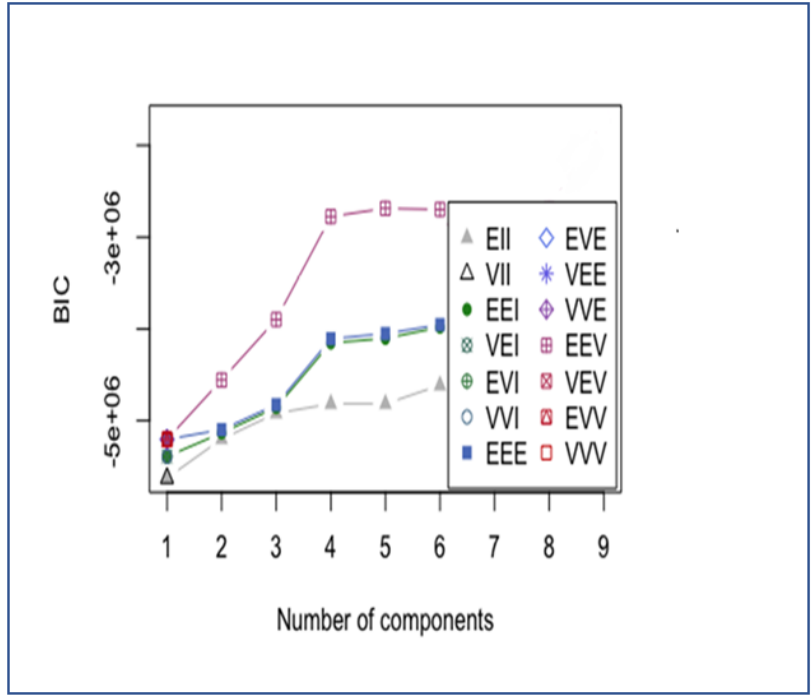

Figure 5 Plot of BIC values for a variety of models and a range of number of clusters

The silhouette score can also measure the goodness of any clustering technique (Alotaik et al, 2017). In the silhouette score, there is a term called $s(i)$ which is calculated as follows:

$$s(i) = \frac{b(i) - a(i)}{max\{a(i),\, b(i)\}}$$

Where $a(i)$ is the average dissimilarity of the data value $i$ (can be any variable in the dataset) with all the other data within the same cluster. $b(i)$ is the lowest average dissimilarity of $i$ to any other cluster. Fig. 6 shows the average silhouette score for different number of clusters.

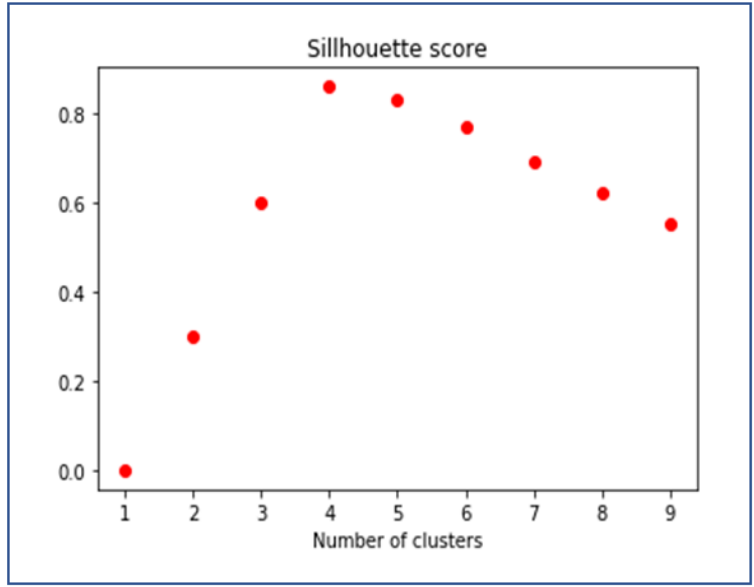

Figure 6 Average silhouette scores for different number of clusters

From Fig. 6 we observe that the average silhouette score is the highest when the number of clusters is 4 and in Fig. 5, we observe that after a steep rise from clusters 3 to 4 it has been relatively steady when the number of clusters recorded 4. Hence, from BIC, model EEV (Ellipsoidal, equal volume, and equal shape) with 4 clusters is taken as best blend.

**Predictive Modelling Framework**

Fig. 7 depicts the study's framework for predictive modeling. The input data preparation is the first step in the study; during this phase, we will add input variables to the food demand dataset consisting of the clustering results from the previous clustering analysis study. We will investigate the effect of GMM clustering findings on the predictive accuracy of the food demand dataset in this way. As a result, we perform predictive modeling on two datasets: one dataset containing the clustering results as an additional input variable and another dataset without details of the clustering results.

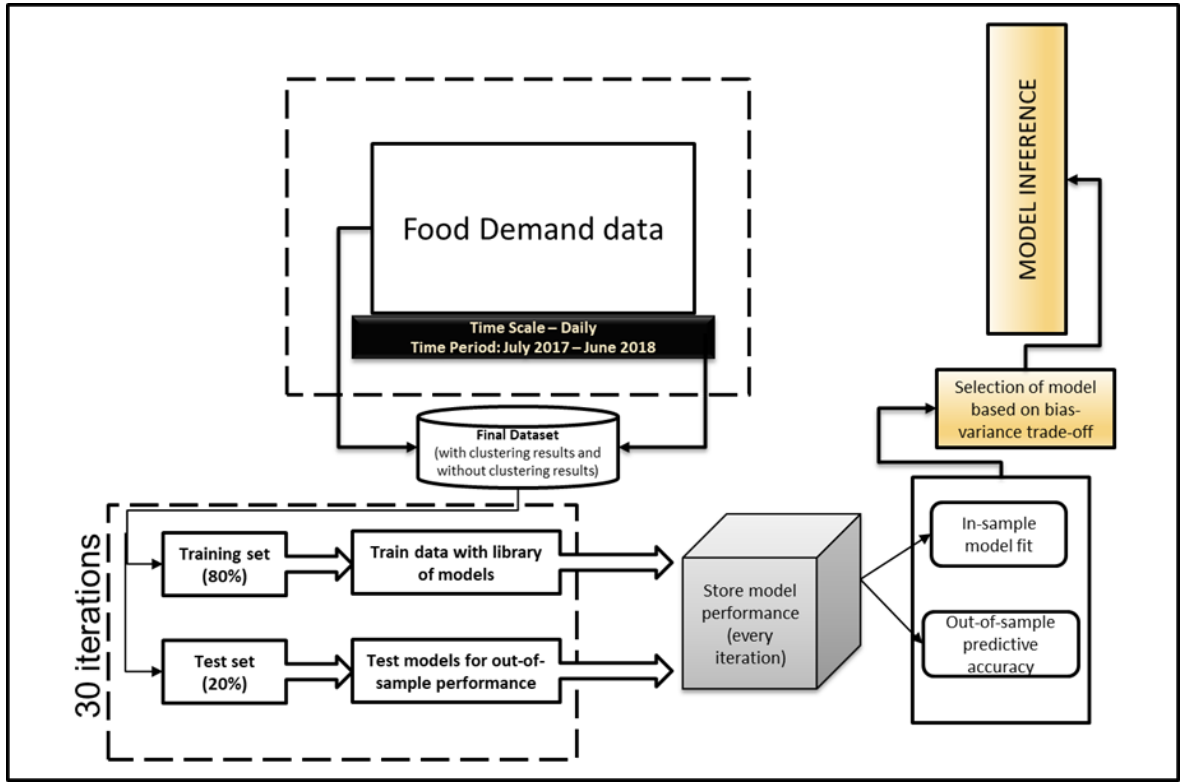

Figure 7 Predictive Modeling Framework

As evident from this Fig. 7, while data specific to the GCFB was used to demonstrate the applicability of the proposed research, the approach and methodology is transferable and can be extended to other food banks and regions.

**Prediction Models**

Numerous types of parametric, semi-parametric and non-parametric machine learning methods have been applied and trained to the food demand dataset (both with and without clustering results). This is done to develop optimum predictive models that portray the best understanding of the complex and non-linear relationships between the demand of food in the food banks and the various input variables. We utilized the methods of Generalized Linear Model (GLM), Generalized Additive Model (GAM), Multivariate Adaptive Regression Splines (MARS), Random Forest (RF), and Bayesian Additive Regression Trees (BART) to estimate the food demand in the given region that GCFB handles. Based on these machine learning algorithms, predictive models of the food demand are developed employing rigorous cross-validation to highlight the model that out-performed all the others in terms of out-of-sample predictive accuracy. A brief review of each of the methods used in our study are examined below.

*Generalized Linear Model (GLM)*

GLM stands for Generalized Linear Models and is an extension of linear regression. The normality assumption is relaxed in GLMs, enabling the response variable to be distributed according to an exponential family of distributions (e.g., Gaussian, Binomial, Poisson, Gamma, or inverse-Gaussian) and linked to the predictors through a link function (Cordeiro & Mccullagh, 1991; McCulloch, 2000). A dependent variable Y with a distribution that falls into the categories of normal, binomial, Poisson, gamma, or Inverse-Gaussian, as shown in the equations below:

$$Y_i \sim f_{Y_i}(y_i)$$

$$f_{Y_i}(y_i) = \exp exp \left\{ \frac{y_i \theta_i - b(\theta_i)}{a(\phi)} + c(y_i, \phi) \right\}$$

where $\theta$ and $\phi$ are the location and scale parameters respectively, a set of independent variables $x_i$, a link function $g(.)$ binding the parameters of the dependent variable to the linear combination of input variables.

*Generalized Additive Model (GAM)*

Generalized Additive Model (GAM) is a semi-parametric machine learning method. It relaxed the assumption of linearity that is considered in the above mentioned GLM method, thereby allocating for regional non-linearities (Hastie & Tibshirani, 1990; Hastie et al., 2009). Here, the dependent variable $y$ has a distribution with mean $\mu = E[Y|x_1, x_2, \ldots x_p]$ (an assumption GAMs makes) associated to the predictor variables via a link function as:

$$g(\mu_i) = \alpha + \sum_{j=1}^{p} f_j(x_j)$$

where each $f_j$ is a smoothing function of a class of functions projected non-parametrically, like regression splines and tensor product splines.

*Multivariate Adaptive Regression Splines (MARS)*

MARS is a semi-parametric, adaptive, and compliant regression technique that is well suited for high-dimensional problems (i.e., datasets with a large number of input variables) (Friedman, 1991). As shown in the equation below, the MARS model uses sum-of-splines to allow the answer to vary non-linearly with the input variables:

$$f(X) = \beta_0 + \sum_{m=1}^{M} \beta_m h_m(X)$$

where each $h_m(x)$ represents the reflected pair of linear splines, , $\beta_0$ represents the intercept and $\beta_m$ represents the vector of the coefficients. $\beta_m$ coefficients are projected by reducing the sum of square errors.

*Random Forest (RF)*

RF is a non-parametric tree-based ensemble data-miner (Breiman, 2001). The procedure consists of $B$ bootstrapped regression trees ($T_b$) with $B$ chosen based on cross-validation. Regression trees are low bias high variance techniques. In other words, they can obtain the shape of the data pretty well (low bias) but are highly vulnerable to outliers (high variance) (Hastie et al., 2009). RF pulls in model averaging as a variance reduction technique. The final estimate is, therefore, the average of predictions across all $B$ trees as shown below:

$$f_{rf}^{B}(x) = \frac{1}{B} \sum_{b=1}^{B} T_b(x)$$

RF can achieve strong predictions by lowering the correlation between the trees such that model averaging can be used to get a low-bias, low-variance predictions, and keeping the errors of every individual unpruned tree low.

*Bayesian Additive Regression Trees (BART)*

Bayesian Additive Regression Trees (BART) is a non-parametric Bayesian method. The final model estimate contains the summation of the estimate from $m$ small trees, as shown in the equation (Merwe, 2009):

$$Y = (\sum_{i=1}^{m} g(x; T_j, M_j)) + \epsilon, \qquad \epsilon \sim N(0, \sigma^2)$$

where $g(x; T, M)$ is the function which designates the parameters of the terminal nodes of tree $T$, $\mu_i \in M$ to the predictors $x$. To ensure that each tree contributes only partially to the final predictions, regularization priors are used to control the model's complexity.

**Bias-variance tradeoff**

The ability of a predictive model to make good predictions on an individual test sample determines its generalization efficiency. The biggest decision maker for minimizing generalization error is balancing the bias-variance trade-off (Hastie et al., 2009). One of the most commonly used approaches for matching bias and variance is cross validation. To approximate predictive precision, we use the k-fold cross validation technique. K-fold cross-validation involves slicing the data into $k$ equal-sized subsets at random. The model is fitted on all subsets except the $k$th held-out sample of each replication, and the predictive accuracy is determined based on the performance of the models on the kth held-out subset. The efficiency of the out-of-sample model was calculated in this paper using a 30% holdout cross validation and the formulae below:

$$MSE_{out-of-sample} = \frac{1}{k} [\sum_{k=1}^{n} \frac{1}{m} (\sum_{i=1}^{m} (y_{i,k} - \hat{y}_{i,k})^2)]$$

$$MAE_{out-of-sample} = \frac{1}{k} \left[ \sum_{k=1}^{n} \frac{1}{m} \mid \sum_{i=1}^{m} \left( y_{i,k} - \hat{y}_{i,k} \right) \mid \right]$$

$k$ = number of times cross-validation was done

$m$ = hold-out numbers during each cross-validation

$y_{i,k}$ = during the *kth* cross-validation, the *ith* actual observation that was kept out at random

$\hat{y}_{i,k}$ = using the model developed, using the training set data during the *kth* cross-validation, and obtaining the predicted *ith* observation

In this paper, we pick models based on both in-sample fit and out-of-sample prediction accuracy. The in-sample MSE (Mean Square Error), MAE (Mean Absolute Error), and adjusted $R^2$ were used to calculate the in-sample error, while the out-of-sample MSE (Mean Square Error) and MAE (Mean Absolute Error) were calculated as previously stated.

**Tuning parameters**

**Generalized Linear Model (GLM):** For GLM, the value of the tuning parameters applied are listed below:

**$k$:** Refers to the <u>number of degrees of freedom used for the penalty</u>. For our study, when k = 2 the best Akaike Information Criterion (AIC) value of 1738437 is obtained: k = log(n). Hence, in our study we used k=2.

***Dist.= Gaussian***: This parameter specifies the <u>type of error distribution and link function</u> to be used in the model. In this study, for all the different methods stated, we assumed that the error follows "Gaussian distribution" and the link function is taken to be an "identity" function.

**Generalized Additive Models (GAM)**: We implemented a stepwise update methodology and cubic smoothing function which generated the best predictive accuracy among the tuning options to select the best fit model.

**Multivariate Adaptive Regression Splines (MARS)**: The tuning parameters used for developing the MARS model are described below:

$pMethod$: It refers to the <u>pruning method</u>. The type of pruning method used is *"cv"*. $pMethod = cv$ uses cross-validation to select the number of terms. This selects the number of terms that gives the maximum mean out-of-fold $R^2$ on the fold models. We selected the model based on the best goodness-of-fit for the models.

$nfold$: This parameter refers to the <u>number of cross-validation folds</u>. In R, default is 0 i.e., no cross validation. If $nfold > 1$, earth first builds a standard model as usual with all the data; then it builds $nfold$ cross-validated models, measuring $R^2$ on the out-of-fold (left out) data each time. The final cross validation $R^2$ ($CVRSq$) is the mean of these out-of-fold $R^2$. The above process of building $nfold$ models is repeated $ncross$ times. In our research, we used the number of cross-validation folds as 10.

$ncross$: This parameter only applies if $nfold > 1$. Each cross-validation has $nfold$ folds. The default in R is 1 and, in our research, we used the value as 5.

**Random Forest (RF):** The tuning parameters for the RF model are considered below.
$mtry$: The <u>number of variables randomly sampled as candidates at each split</u> while growing the trees. To be noted that the default values for the regression tree is $p/3$, where $p$ is the number of independent variables used in the model.
$ntree$: This parameter refers to the <u>number of trees to grow</u>. This must not be set to a very small number to guarantee that every input row will get predicted at least a few rounds. In our research, we selected the value of the $ntree$ that yielded the least mean square error ($mse$) while growing the trees.

**Bayesian Additive Regression Trees (BART):** The tuning parameters used in the BART model are described below:
$k$: For regression, $k$ <u>determines the prior probability</u> that $E(Y|X)$ is contained in the interval $(y\_\{min\}, y\_\{max\})$, based on a normal distribution. For example, when we have $k = 2$, we get the prior probability to be 95%. For classification, k determines the prior $E(Y|X)$ between $(- 3, 3)$. Note that a larger value of k results in more shrinkage and a more conservative fit.
$nu$: It refers to the <u>degrees of freedom</u> for the inverse $\chi^2$ prior.
$q$: This parameter refers to the <u>quantile of the prior on the error variance at which the data-based estimate is assigned</u>. It is to be stated that greater the value of q, the more forceful is the fit; this is because it corresponds to placing more prior weight on values lower than the data-based estimation
. It is not used for classification.
$m$: This parameter refers to the <u>number of trees to be grown</u> in the sum-of-trees model.

## 4. Results and Discussion

**The results of GMM clustering**

The GMM parameterized by the EM algorithm is applied to the aggregated dataset containing various socio-economic and demographic details of the region supported by GCFB. As shown in Table 5, the data is divided into 4 clusters. The four clusters have been named based on the proximity of the families with their respective food agencies. The average distance between clusters 1 and 2 is 1.03 miles, between 2 and 3 is 3.11 miles, and between 3 and 4 is 14.65 miles. These clusters are reasonable considering that we intend to find factors concerning food insecurity for an urban area from the dataset.

**Summarization of clustering patterns**

In Table 5, we observe that the further away the families are from the food agencies, the more people there are in the family. This holds for children and especially adults, but not for seniors. From this table, we can interpret the number of children and seniors having low access to food resources. Around 13,967 children live very far away (19.47 miles on average) from the food agencies, while 122884 children live less than 0.42 miles away from the nearest food assistance.

Also, the number of tracts increases as the distances from the food assistance increases. This makes sense since they are more scattered over the urban areas. Regarding the income of families in each cluster, we can see that most families (89.1%) live in tracts that are considered poor (It is considered as a poor tract if average household income is less than Ohio's median income, according to (Mattogno et al., 2014)). Additionally, around 11% of the families live in tracts with high-income levels.

Table 5 Detailed calculation results of GMM clustering

| Variables | Cluster 1 - proximate | Cluster 2 – reasonable distance | Cluster 3 - Distant | Cluster 4 – extremely distant | Total |
|---|---|---|---|---|---|
| Number of families | 197,844 | 199,475 | 195,081 | 17,009 | 609,409 |
| Number of adults | 221,523 | 244,783 | 245,107 | 23,431 | 734,844 |
| Number of children | 122,884 | 154,202 | 152,667 | 13,967 | 443,720 |
| Number of seniors | 120,513 | 127,827 | 130,792 | 10,417 | 389,549 |
| Number of people | 464,920 | 526,812 | 528,566 | 47,815 | 1,568,113 |
| Average number of adults in family | 1.12 | 1.23 | 1.26 | 1.38 | 1.21 |
| Average number of children in family | 0.62 | 0.77 | 0.78 | 0.82 | 0.73 |
| Average number of seniors in family | 0.61 | 0.64 | 0.67 | 0.61 | 0.64 |
| Average number of people in family | 2.35 | 2.64 | 2.71 | 2.81 | 2.57 |
| Number of tracts | 464 | 545 | 654 | 884 | 943 |
| Average Distance (miles) | 0.42 | 1.45 | 4.63 | 19.47 | 2.64 |
| Coverage | 32.5% | 32.7% | 32.0% | 2.8% | 100.0% |
| Pct of Poor People | 92.8% | 91.2% | 84.3% | 74.7% | 89.1% |
| Pct of Rich People | 7.2% | 8.8% | 15.7% | 25.2% | 10.9% |

The clustering has contributed to discovering that the farther families are from food, the more likely they are to live in a tract with a high-income population. The coverage rate is the number of families within a mile that are served by at least one food bank. The research by Alotaik et al. (2017) is the only one we know that did a similar analysis. However, they introduced census tracts, which are more common than the family's locations. Taking the same assumption used by Alotaik et al. where they considered the possibility of moving food agencies, it can be seen that to increase the coverage of supply of food assistance to the poor people located very far away, some of the food agencies should be moved from cluster 1 to cluster 4 where there are areas with less coverage and people with low income. We do not present the specific tracts and locations of families considering the vast amount of data provided for this region, but they are very well known, and we can quickly identify can be easily in the given dataset.

Fig. 8 shows the spread of the agencies and families in each cluster. We use the latitude and longitude information to plot the corresponding graph for each cluster on the map of the Ohio region. It is visible that the distance increases in each cluster with the average distances of each cluster mentioned in Table 5.

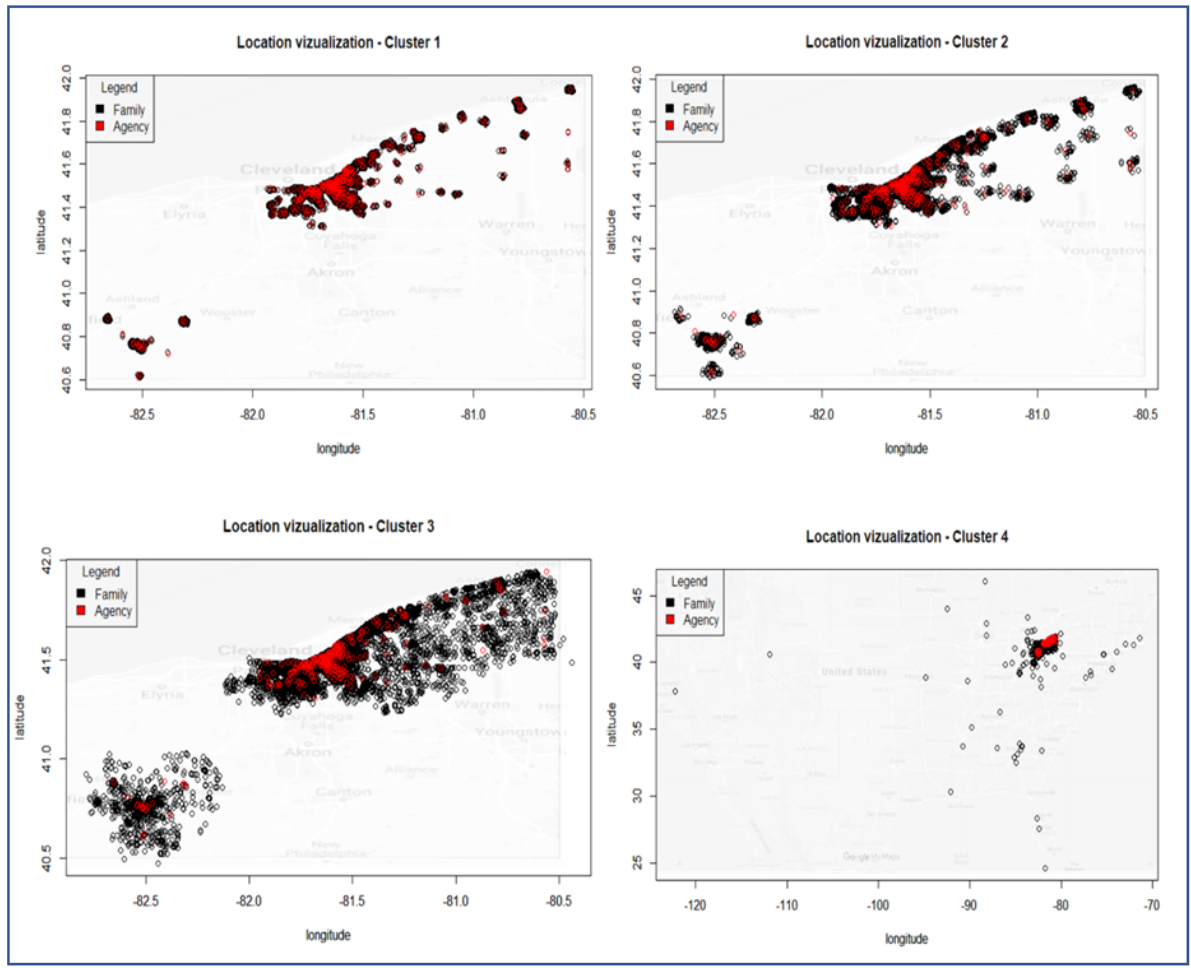

Figure 8 Distance between families and agencies in each cluster

**Modeling of donated food demand**

We developed predictive models for the food demand study considering clustering results information as another independent variable in the food demand dataset (Table 2) and without clustering results to observe any accuracy improvement in the prediction results. We trained the food demand dataset (with and without clustering results separately) with the methods of Generalized Linear Model (GLM), Generalized Additive Model (GAM), Multivariate Adaptive Regression Splines (MARS), Random Forest (RF), and Bayesian Additive Regression Trees (BART). In this section, we will discuss the performance of each of the trained models and choose the one with the final model based on the one that has the best out-of-sample predictive accuracy. Tables 6 and 7 summarize the goodness-of-fit and predictive performance of each trained model. The percentage improvement yielded by each trained model over having no statistical model and using the historical average as a predictor (i.e., the 'mean-only' model) is provided in Tables 8 and 9.

We see that BART substantially outperforms all the other models in terms of goodness-of-fit. Comparing our results of the GLM with the results of BART supports our hypothesis that linear models do not adequately capture the complex non-linearities in food demand data.

Fig. 9 provides the plots of predicted versus observed food demands for the data included with clustering results, and Fig. 10 provides the data excluding the clustering results. In the case of the former, the 95 % credible intervals provide 57.31% coverage for all the observations, whereas the 95% prediction interval offers a 97.68% coverage (Fig. 9). In the case of the data without clustering information, the 95% credible intervals provide 20.65% coverage for all the observations, whereas the 95% prediction interval offers a 95.98% coverage (Fig. 10.).

By observing the results, we conclude that although BART does provide the best predictive accuracy for both datasets, the dataset without clustering results has an unsatisfactory overall error level. As shown in Fig. 9, the results of the models for the dataset consisting of clustering results have been greatly improved. However, the models tend to underestimate the more extreme ends of demand.

Table 6 Modelling with clustering results

| Model | Tuning Parameters | $R^2$ | In-sample | | Out-of-sample | |
|---|---|---|---|---|---|---|
| | | | RMSE | MAE | RMSE | MAE |
| Mean (Null model) | - | - | - | - | 443.12 | 328.37 |
| Generalized Linear Model (GLM) | k=2.0, Dist.=Gaussian | 0.55 | 233.65 | 189.82 | 247.21 | 193.92 |
| Generalized Additive Model (GAM) | Stepwise update | 0.58 | 199.23 | 174.63 | 213.51 | 183.26 |
| Multivariate Adaptive Regression Splines (MARS) | pMethod: cv; nfold: 10; ncross=5 | 0.42 | 217.72 | 195.27 | 248.56 | 199.33 |
| Random Forest (RF) | mtry=p/3 =3; ntree=100 | 0.63 | 142.82 | 134.19 | 155.94 | 140.32 |
| **Bayesian Additive Regression Trees (BART)** | **k=2,nu=10,q=0.75, m=50** | **0.82** | **137.17** | **99.31** | **143.57** | **105.11** |

Table 7 Modelling without clustering results

| Model | Tuning Parameters | $R^2$ | In-sample | | Out-of-sample | |
|---|---|---|---|---|---|---|
| | | | RMSE | MAE | RMSE | MAE |
| Mean (Null model) | - | - | - | - | 683.34 | 510.39 |
| Generalized Linear Model (GLM) | k=2.0, Dist.=Gaussian | 0.28 | 379.36 | 354.49 | 399.53 | 357.51 |
| Generalized Additive Model (GAM) | Stepwise update | 0.31 | 363.69 | 335.62 | 378.57 | 329.26 |
| Multivariate Adaptive Regression Splines (MARS) | pMethod: cv; nfold: 10; ncross=5 | 0.35 | 312.23 | 289.60 | 340.41 | 299.46 |
| Random Forest (RF) | mtry=p/3 =3; ntree=100 | 0.41 | 278.33 | 241.91 | 281.26 | 263.51 |
| **Bayesian Additive Regression Trees (BART)** | **k=2,nu=10,q=0.75,m=50** | **0.47** | **226.61** | **202.11** | **235.36** | **216.44** |

Table 8 Percentage improvement over the 'null' model for modelling with clustering results

| Models | Out-of-sample error (%imp) | |
|---|---|---|
| | RMSE | MAE |
| GLM | 44.2 | 40.9 |
| GAM | 51.8 | 44.2 |
| MARS | 43.9 | 39.3 |
| RF | 64.8 | 57.2 |
| **BART** | **67.6** | **67.9** |

| Models | Out-of-sample error (%imp) | |
|---|---|---|
| | RMSE | MAE |
| GLM | 41.5 | 29.9 |
| GAM | 44.6 | 35.4 |

| MARS | 50.2 | 41.3 |
| RF | 58.8 | 48.3 |
| **BART** | **65.5** | **57.5** |

Table 9 Percentage improvement over the 'null' model for modelling without clustering results

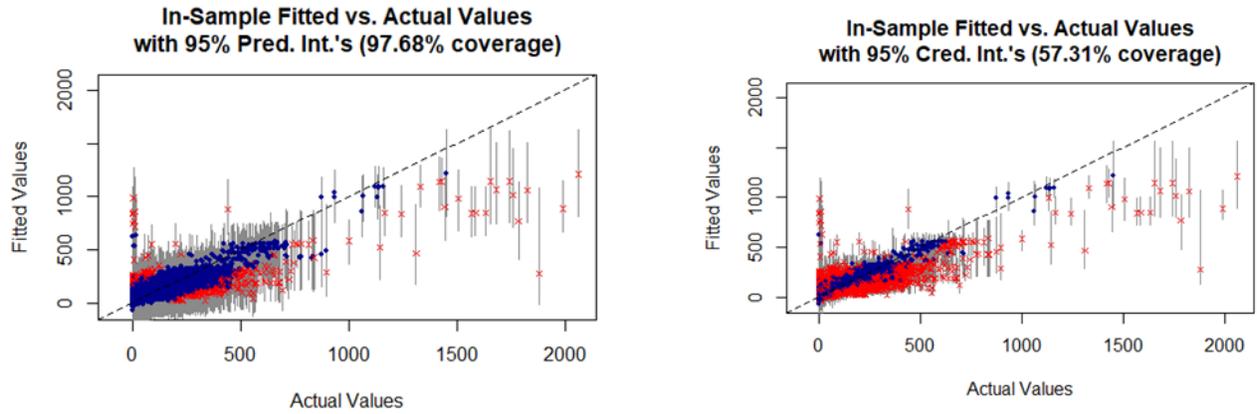

Figure 9 The prediction results of food demand dataset with clustering results

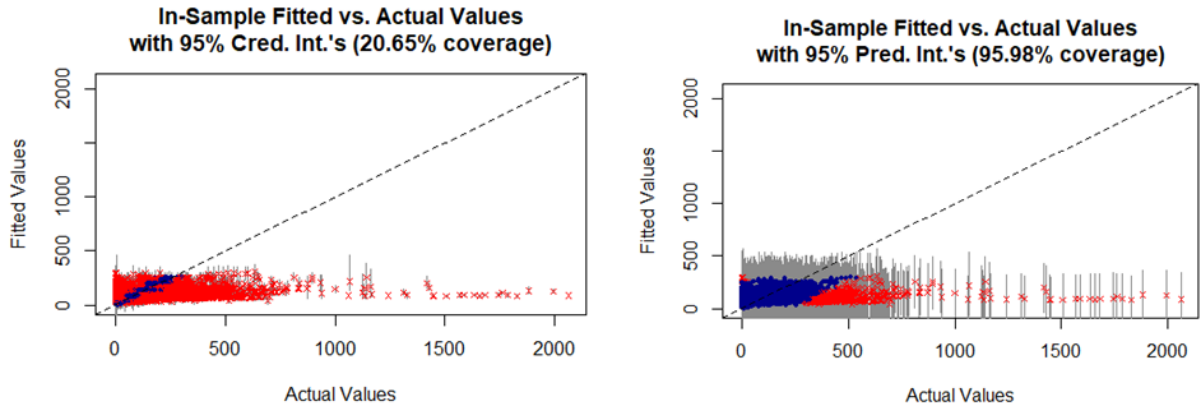

Figure 10 The prediction results of food demand dataset without clustering results

## 5. Discussion

Observing the results in Section 4 shows how clustering results in predictive modeling of food demand aids in better forecasting accuracy. This leads us to implement a streamlined research framework to be used by the food bank officials. The research framework proposed is the two-stage hybrid demand estimation model to identify and classify the aggregated dataset to clusters and use the cluster results on the food demand dataset (obtained from the aggregated dataset) for predictive modeling by Bayesian Additive Regression Trees (BART). The outcome of this framework is to understand and aid the food bank management with the food demand behavior with greater accuracy and optimal planning.

Fig. 11 depicts the proposed approach for developing the data-driven demand estimation model. The significant steps include data collection and data wrangling, followed by implementing algorithm-based statistical learning methods for classification and prediction.

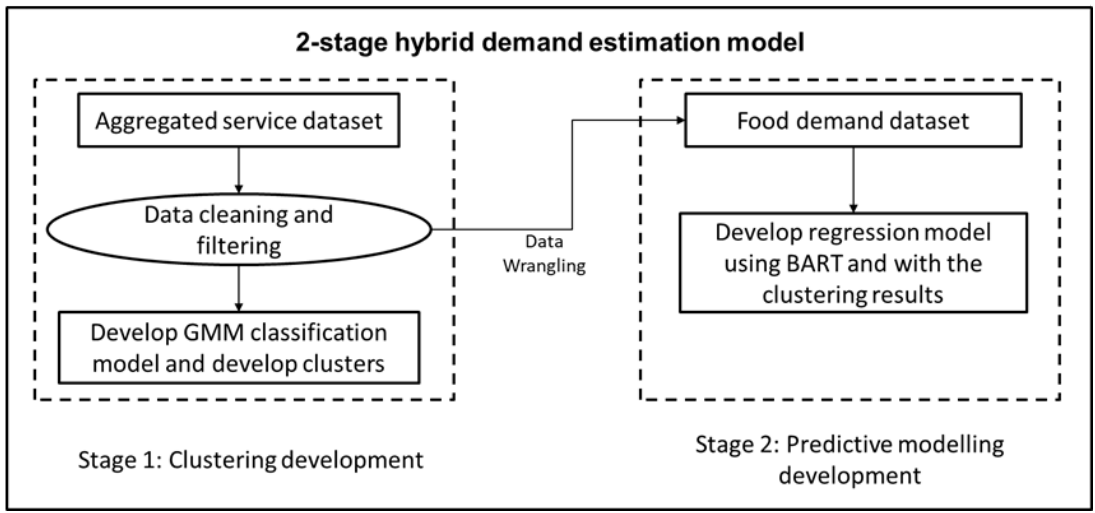

Figure 11 Flow chart of research methodology steps for developing two-stage hybrid demand estimation model

**5.1 Proposed predictive framework implementation on use case studies: an observation**

Our research framework of implementing predictive modeling in the food bank logistics can be used across different planning projects and situations such as budgeting, facility location, routing, cost reduction, and optimization. The predictive analysis aids in every stage of planning. In order to validate the applicability of the proposed framework for each stage of planning, we provide its implementation in two case studies: Warehouse costing and improvement (tactical planning case) and Budget planning of a food bank organization (strategic planning case). Case 1 analyzes the procedure for budget planning, while case 2 analyzes the space planning and costing of a warehouse owned by a food bank. We provide the implementation of the results of the proposed framework onto these planning problems and present predicted results in each of the case studies to demonstrate the potential of the proposed framework and its straightforward implementation in achieving optimal planning performance.

**5.1.1 Case 1: Strategic Level Planning- Predictive budgeting for food banking**

Effective management of finances is critical to organizational success. Budgeting will outline the high costs and give an overview of available capital. Monetary Donors also find tracking their contributions helpful to see how the food banks use their funds. Having a wide-ranging budget will establish integrity with the donors and provide a clear view of setting the goals for the following year (Gutjahr & Fischer, 2018). We use the strategic planning report of a food bank in literature as a template (Second Harvest Food Bank of Central Florida, 2020). In the report, the budgets are drawn to cover a fiscal year and must be made ready before the beginning of each year. To set reasonable projections of financial need, it is essential to have accurate forecasts and analyses. Based on the needs of budgeting, the proposed two-stage hybrid demand estimation model is implemented on our dataset to provide the current food demand based on the total number of people and type of people (adults, children, and seniors) for displaying the accuracy of the forecasts and finally provides a forecast of the coming year count of people. Since food banks develop the budgets annually, Table 10 provides the predictive results on an annual basis.

Table 10. Demand (count of people) from food agencies on an annual basis

| People count | Count in 2018 (Actual) | Count in 2018 (Forecasted) | Variance (%) | Count in 2019 (Forecasted) |
|---|---|---|---|---|
| Adults | 734,844 | 765707 | 4.2 | 783753 |

| | | | | | |
|---|---|---|---|---|---|
| Children | 443,720 | 458806 | 3.4 | 488234 | |
| Seniors | 389,549 | 402014 | 3.2 | 412023 | |

From the results mentioned in Table 10, we can see that the variance of the predicted values to the actual values is less than 5% which is beneficial in the budget planning as it is recommended to ensure variance is less than 10% for efficient budgeting (Shim, Siegel, & Shim, 2012).

### 5.1.2 Case 2: Tactical Level Planning – Warehouse costing and improvement of a food bank system

As mentioned in Section 1, food banks distribute and provide food to their respective food agencies with the food insecure people receiving food by visiting these food agencies. Therefore, the food banks need to ensure optimal warehouse spacing and inventory management to maintain the food products donated and handle the growing food distribution and demand (Shrestha, 2009). The usage of the proposed two-stage hybrid demand estimation model ensures the success of accomplishing this challenge faced by the tactical planners of the food bank system. Using the proposed framework, we can determine the peak months of demand and visually provide aid to the management team, understand their load of inventories, and prepare for handling them in the future using the given dataset. Table 11 below provides the actual demand in terms of the count of people, and we provide the results for prediction extracted monthly for the upcoming period of 2019 in Table 12.

Table 11. Monthly demand (count of people) from food agencies (2018)

| People count | January | February | March | April | May | June | July | August | September | October | November | December |
|---|---|---|---|---|---|---|---|---|---|---|---|---|
| Adult | 43168 | 36007 | 25467 | 38720 | 26957 | 21514 | 35280 | 78946 | 236309 | 103770 | 86028 | 44168 |
| Children | 28885 | 21917 | 15703 | 23576 | 16469 | 13796 | 21475 | 48054 | 143840 | 63164 | 52365 | 26885 |
| Seniors | 39964 | 30352 | 14581 | 21890 | 15237 | 12160 | 19941 | 44668 | 133566 | 58653 | 48625 | 24964 |

Table 12. Forecasted monthly demand (count of people) from food agencies for 2019

| People count | January | February | March | April | May | June | July | August | September | October | November | December |
|---|---|---|---|---|---|---|---|---|---|---|---|---|
| Adult | 43486 | 39406 | 39898 | 50674 | 45464 | 45529 | 75419 | 88788 | 196924 | 86475 | 71690 | 40486 |
| Children | 37644 | 23890 | 24286 | 30845 | 27674 | 27713 | 45896 | 54045 | 119867 | 52637 | 43637 | 24644 |

| Seniors | 12884 | 12160 | 21550 | 20502 | 25690 | 25733 | 42617 | 50185 | 111305 | 48877 | 40520 | 22884 |

## 6. Conclusions

This paper proposes the characteristics of a particular region using a clustering method such as GMM in terms of accessibility of food assistance and finding ways to increase their access to food. With this information, GCFB can manage and distribute their food resources to the food insecure efficiently and equitably by targeting the regions of food assistance desserts, increasing the coverage in regions of people receiving low income and located far away from the source of food assistance. By the results of the food accessibility pattern study, we study the food demand of the GCFB organization by developing predictive models. We see that implementing clustering results to the predictive models has a noticeable accuracy improvement; hence, a two-stage hybrid demand estimation model is proposed based on the results obtained. A future direction in this research is using these predictive model results of food demand in a given region as input parameters to mathematical models developed to improve the equitable distribution of donated foods to the people in need.

## Acknowledgements

We would like to express our sincere appreciation to Phil Trimble and the team from the Greater Cleveland Food Bank for their valuable input into this project.